\documentclass[nofootinbib,a4paper,aps,prd,10pt,superscriptaddress,showkeys,showpacs,twocolumn]{revtex4}

\usepackage{graphicx}
\usepackage{amsfonts}
\usepackage{amssymb} 
\usepackage{amsmath}
\usepackage{hyperref}
\usepackage{natbib}
\usepackage{float}
\usepackage{dcolumn}
\usepackage{bm}
\usepackage{latexsym,color}

\def\nat{Nature}
\def\prl{Phys. Rev. Lett.}

\def\prd{Phys. Rev. D}

\def\mnras{Mon. Not. Roy. Astr. Soc.}
\def\apj{Astrophys. J.}
\def\apjl{Astrophys. J. Lett.}

\def\aap{Astron. Astrophys.}

\def\araa{Annual Rev. of Astron. Astrophys.}

\newcommand{\bea}{\begin{eqnarray}}
\newcommand{\eea}{\end{eqnarray}}
\newcommand{\be}{\begin{equation}}
\newcommand{\ee}{\end{equation}}

\begin{document}

\title{Luminosity of accretion disks in compact objects with quadrupole}

\author{Kuantay~\surname{Boshkayev}}
\email[]{kuantay@mail.ru}
\affiliation{National Nanotechnology Laboratory of Open Type,  Almaty 050040, Kazakhstan.}
\affiliation{%
Al-Farabi Kazakh National University, Al-Farabi av. 71, 050040 Almaty, Kazakhstan.
}

\author{Talgar~\surname{Konysbayev}}
\email[] {talgar\_777@mail.ru}
\affiliation{National Nanotechnology Laboratory of Open Type,  Almaty 050040, Kazakhstan.}

\affiliation{%
Al-Farabi Kazakh National University, Al-Farabi av. 71, 050040 Almaty, Kazakhstan.
}

\author{Ergali~\surname{Kurmanov}}
\email[]{kurmanov.yergali@kaznu.kz}
\affiliation{National Nanotechnology Laboratory of Open Type,  Almaty 050040, Kazakhstan.}

\affiliation{%
Al-Farabi Kazakh National University, Al-Farabi av. 71, 050040 Almaty, Kazakhstan.
}

\author{Orlando~\surname{Luongo}}
\email[]{orlando.luongo@unicam.it}
\affiliation{National Nanotechnology Laboratory of Open Type,  Almaty 050040, Kazakhstan.}

\affiliation{%
Al-Farabi Kazakh National University, Al-Farabi av. 71, 050040 Almaty, Kazakhstan.
}

\affiliation{%
Dipartimento di Matematica, Universit\`a di Pisa, Largo B. Pontecorvo 5, Pisa, 56127, Italy.
}
\affiliation{%
Divisione di Fisica, Universit\`a di Camerino, Via Madonna delle Carceri, 9, 62032, Italy.
}

\author{Daniele~\surname{Malafarina}}
\email[]{daniele.malafarina@nu.edu.kz}
\affiliation{%
Department of Physics, Nazarbayev University, Kabanbay Batyr 53, 010000 Nur-Sultan, Kazakhstan.
}

\author{Hernando~\surname{Quevedo}}
\email[]{quevedo@nucleares.unam.mx}
\affiliation{%
Al-Farabi Kazakh National University, Al-Farabi av. 71, 050040 Almaty, Kazakhstan.
}
\affiliation{%
Instituto de Ciencias Nucleares, Universidad Nacional Aut\`onoma de M\`exico, Mexico. 
}
\affiliation{%
Dipartimento di Fisica and ICRA, Universit\`a di Roma “La Sapienza”, Roma, Italy.
}

\date{\today}

\begin{abstract}
We consider the circular motion of test particles in the gravitational field of a static and axially-symmetric compact object described by the $q$-metric. To this end, we calculate orbital parameters of test particles on accretion disks such as angular velocity ($\Omega$), total energy ($E$), angular momentum ($L$), and radius of the innermost stable circular orbit ($r_{ISCO}$) as functions of the mass ($m$) and quadrupole ($q$) parameters of the source. 
The radiative flux, differential, and spectral luminosity of the accretion disk, which are quantities that can be experimentally measured, are then explored in detail. The obtained results are compared with the corresponding ones for the Schwarzschild and Kerr black holes in order to establish whether black holes may be distinguished from the $q$-metric via observations of the accretion disk's spectrum.
\end{abstract}



\maketitle
 
\section{Introduction}

The spectra of accretion disks surrounding massive compact objects are routinely observed in astrophysics. Such observations provide invaluable information about the nature of the accreting central object and for extremely condensed sources one must necessarily use general relativity to describe both the source as well as the accreting matter. The accretion disk luminosity of compact objects can be modeled for many solutions to Einstein's field equations. 
In particular, these solutions may describe the gravitational field of neutral black holes \cite{2012LRR....15....7C} or they may describe the field outside of massive compact objects, some realistic like white dwarfs and neutron stars, and others more exotic, such as for example boson stars \cite{cardopan,beheshagasp} or gravastars \cite{2004PNAS..101.9545M}. 

As of now, we can not rule out the possible existence of exotic compact objects in the universe as most observations of black hole candidates are still not able to probe the geometry in the vicinity of such astrophysical sources. Such observations include gravitational waves emitted from the inspiral of binary black holes \cite{2016PhRvL.116f1102A,2016PhRvL.116v1101A}, the motion of stars near the galactic center \cite{2018A&A...615L..15G, 1998ApJ...509..678G, 2000Natur.407..349G}, the shadow of the supermassive black hole candidate at the core of the galaxy M87 \cite{2019ApJ...875L...1E}, spectra from x-ray binary systems \cite{x-ray}, and light spectra from accretion disks surrounding supermassive black hole candidates at the core of galaxies \cite{agn1, agn2}. 
The last two sets of observations are the most abundant ones and both rely on the measurement of the accretion disk's spectra. As it is well known, some features of the accretion disk depend on the space-time geometry and therefore can in principle be used to obtain constraints on the geometry from observations \cite{abramowicz}.

Generally speaking, relativistic compact objects are modeled by assuming a space-time metric fulfilling given conditions of symmetry for the exterior and solving the equations for hydrodynamic equilibrium for the matter content in the interior. 
Often ``exotic'' matter can be used to describe hypothetical objects that are massive, compact, and stable. This speculation has proven helpful to provide either black hole alternatives or dark matter candidates, also leading to the suggestion that such massive objects may be dark matter condensates \cite{2020MNRAS.496.1115B, Ruffini}\footnote{For a different perspective on the nature of dark matter and dark energy see e.g.~\cite{LM}.}.

Nevertheless, we know that general relativity is an incomplete theory and therefore the classical description of the geometry and/or the matter models may fail when the field becomes sufficiently strong \cite{2016PhRvL.116c1101J, 2016Univ....2....7B, 2016CQGra..33w5010G, malafarina2017}. The limits of the theory are signalled by the appearance of singularities. For example, vacuum, static, and axially symmetric solutions (known as Weyl's class \cite{1917AnP...359..117W, 1919AnP...364..185W}) generically exhibit curvature singularities at the infinitely redshifted surface which corresponds to the horizon in the black hole case \cite{Quevedo1990, Pastora1994}. Such solutions may be considered as describing the exterior field of static and axially symmetric exotic compact objects whose boundary is located at a distance outside the singularity.

In this paper, we consider one of such solutions describing a static deformed compact object, the so-called $q$-metric \cite{1966JMP.....7.1137Z, PhysRevD.2.2119}. 
In the literature, the metric is also known as the Zipoy-Voorhees metric, $\delta$-metric and $\gamma$-metric. Despite the different nomenclature, it is practically the same metric, which here we prefer to denote with $q$-metric to stress the importance of the quadrupole parameter $q$ \cite{2016PhRvD..93b4024B}.

The line element of the $q$-metric is particularly suitable to study the exterior field of exotic compact objects and its relation to the Schwarzschild space-time because it involves only two parameters of clear physical interpretation: $m$ which is related to the mass of the source and $q$ which relates to the quadrupole and hence describes the departure of the object from spherical symmetry \cite{ZV3}.

It is worth noticing that relativistic matter sources for metrics of Weyl's class do exist \cite{hernandez, bonnor}. In particular some early attempts to develop interiors for the $q$-metric were made in \cite{1982GReGr..14...97S, HMM}, and, more recently, a complete solution has been proposed in \cite{hphm16} and an approximate solution in \cite{abishevetal19}, which satisfy all the conditions to be considered physically meaningful.
Furthermore, several studies of the physical features of the $q$-metric can be found in the literature, e.g. shadows around the $q$-metric \cite{2021CQGra..38a5008A,2021arXiv210512057S, shadow}, geodesics \cite{2012PhRvD..85j4031C, 2016PhRvD..93b4024B,2020arXiv201015723F}, quasinormal modes \cite{2019PhRvD..99d4005A, charge1}, motion of charged particles \cite{charge2} and spinning particles \cite{spin}, neutrino oscillations \cite{2020EPJC...80..964B}, particle's collisions \cite{malafarina2021}, etc.

However, in order to study how the features of the $q$-metric deviate from those of a black hole geometry, it is important to consider quantities that can be measured via observations. Assuming that the compact source is surrounded by an accretion disk, the direct measurement of the innermost stable circular orbit (ISCO) of the disc is practically impossible. On the other hand, the accretion spectrum of the disk, which does depend on the ISCO, is a quantity that can be observed with modern telescopes.
Hence, we here investigate the luminosity of thin accretion disks in the space-time described by the $q$-metric employing the widely-consolidated model proposed by Novikov and Thorne, Page and Thorne \cite{novikov1973, page1974}. In so doing, we first calculate the orbital parameters of test particles and then numerically construct the simulated radiative flux, spectral and differential luminosities of the disk as functions of the radial distance from the source and the frequency of the emitted radiation. 
In order to establish whether observations of accretion disk spectra would be able to constraint the value of the mass quadrupole, we then compare the results obtained for the $q$-metric with the ones for the Kerr metric. 
It is worth noticing that models of accretion disks' spectra are commonly used to investigate the observational properties of hypothetical compact objects (see for example \cite{KH, JMN, 2020MNRAS.496.1115B}).

The paper is organized as follows. In Sect. \ref{sez2}, we describe the main features of the $q$-metric and briefly review the formalism for the motion of test particles on accretion disks, while in Sect. \ref{sez2b} the same discussion is briefly reviewed for the Kerr space-time. In Sect. \ref{sez3}, we review the thin accretion disks formalism using the Novikov-Thorne and Page-Thorne models and in Sect. \ref{sez4} we apply it to the $q$-metric and compare the results with the corresponding ones for the Kerr space-time. Finally, in Sect. \ref{sez5} we present the  conclusions and perspectives of our work.
Throughout the paper we make use of geometrized units setting $G=c=1$.


\section{Particle motion in the {\it q}-metric}\label{sez2}

It is well-known that the most general axisymmetric and static solution of the vacuum field equations is represented by the Weyl line element \cite{2003esef.book.....S} which describes an infinite number of solutions. Among them the so-called $q$-metric is considered to be the simplest generalization of the Schwarzschild metric containing a quadrupole parameter
\cite{Quevedo, fqs18}. Other solutions that generalize the Schwarzschild metric to include a quadrupole moment have also been proposed \cite{2019PhRvD.100l4021M, ER, GM}.
Here we focus on the $q$-metric which is given by the line element

\begin{eqnarray}
ds^2&=&\left(1-\frac{2m}{r}\right)^{1+q}dt^2-\left(1-\frac{2m}{r}\right)^{-q}\nonumber\\
&\times&\Bigg[ \left(1+\frac{m^{2}\sin^{2}\theta}{r^{2}-2mr}\right)^{-q(2+q)}\left(\frac{dr^{2}}{1-\frac{2m}{r}}+r^{2}d\theta^{2}\right)\nonumber\\
&+&r^{2}\sin^{2}\theta d\phi^{2}\Bigg],\
\label{eq:metric}
\end{eqnarray}
where $m$ and $q$ are the mass and dimensionless quadrupole parameters respectively and the allowed values for $q$ are $1+q>0$. It is easy to notice that for vanishing $q$ one recovers the Schwarzschild metric while values of $q>0$ ($q<0$) describe the exterior field of an oblate (prolate) source. Also it is well known that the $q$-metric exhibits a curvature singularity at $r=2m$ for all values of $q\neq 0$ implying that the matter source must have a boundary $r_b>2m$.

Furthermore the total gravitational mass of the source, as measured by faraway observers, can be easily obtained, for example from the evaluation of Komar's integral, and it is given by $M_T=m(1+q)$.
It is therefore interesting to model the exterior field of compact objects with this metric and to check the effects that the quadrupole parameter $q$ has on the observable features such as accretion disks. To do so, we need to first compute circular geodesics and the innermost stable circular orbit (ISCO).   

\subsection{Circular orbits in the {\it q}-metric}

We limit our attention to the equatorial plane, where  $\theta=\pi/2$ and consider particles on circular orbits. The main orbital quantities for particles on a circular orbit at a radius $r$ are then given by 
\bea 
\Omega^{2}&=&\left(1-\frac{2m}{r}\right)^{1+2q}\frac{(1+q)m}{r^{2}(r-(2+q)m)},\label{eq:omega} \\
E^{2}&=&\left(1-\frac{2m}{r}\right)^{1+q}\frac{r-(2+q)m}{r-(3+2q)m}, \label{eq:energy} \\ 
L^{2}&=&\left(1-\frac{2m}{r}\right)^{-q}\frac{(1+q)mr^{2}}{r-(3+2q)m}, \label{eq:angmom} 
\eea
where $\Omega=\Omega(r)$ is the orbital angular velocity, $E=E(r)$ is the energy per unit mass and $L=L(r)$ is the orbital angular momentum per unit mass of the test particle. As stressed above, these quantities reduce to the corresponding values for the Schwarzschild metric as $q\to 0$.
The orbital parameters for particles in the $q$-metric have been studied by several authors (see for example \cite{HPS, 2012PhRvD..85j4031C}).

The ISCO is the closest stable circular orbit that massive test particles can have around a compact object. In astrophysics it usually determines the inner edge of the accretion disk and its value depends on the parameters characterizing the source, namely the mass, angular momentum and higher mass multipoles. Its role in investigating astrophysical compact objects turns out to be essential since, by observing accretion disks, one can obtain an estimate of the ISCO and consequently constrain the values of the corresponding parameters for the source.
For example, for a Kerr black hole, if an independent measure of the mass is available, then a measurement of the ISCO allows to estimate the value of the angular momentum
(in practice things are more complicated, see for example \cite{mcclintock}). 
Similarly for the $q$-metric, in principle, knowing the active gravitational mass of the source, by measuring the ISCO one could obtain an estimate of the mass quadrupole moment.
It is important to notice that different geometries may mimic each other. For example, a static solution such as the $q$-metric with a given value of $q$ may produce the same ISCO as the Kerr metric for a given value of the angular momentum thus requiring more than one measurement in order to be able to distinguish the two geometries. 

The radius of the innermost stable circular orbit $r_{ISCO}$ is defined via the condition $dL/dr=0$ \cite{2016PhRvD..93b4024B}. So, for the $q$-metric, taking Eq. \eqref{eq:angmom}, we get the quadratic equation 

\begin{equation}
 \label{eq:risco0}
r^2-2m\left(4+3q\right)r+2m^2\left(6+7q+2q^2\right)=0,  
\end{equation}
 and thus two solutions for the ISCO given by
\begin{equation}
 \label{eq:risco}
r_{ISCO}^{\pm}=m\left(4+3q\pm \sqrt{5q^2+10q+4}\right),  
\end{equation}
where $\pm$ signs indicate the inner and outer radii, respectively. The solution $r_{ISCO}^+$ is the physical one corresponding to the edge of the accretion disk and it is the solution that we will consider hereafter. This solution exists for $q\geq 1/\sqrt{5}-1$. The inner solution $r_{ISCO}^-$ is not physical since for most values of $q$ it is situated either below the singularity, i.e. $r_{ISCO}^-<2m$ or below the photon capture radius, i.e. $r_{ISCO}^-<m(3+2q)$. It is nevertheless worth noticing that there exists a range of values of $q\in[1/\sqrt{5}-1,-1/2]$ for which the inner ISCO may be considered physical. In this case there exist two separate regions outside the singularity where stable circular orbits are allowed, one for $r>r_{ISCO}^+$ and one for $r<r_{ISCO}^-$.
For $q=1/\sqrt{5}-1$ the two solutions for the ISCO radius coincide $r_{ISCO}^+=r_{ISCO}^-$ while for $q<1/\sqrt{5}-1$ equation \eqref{eq:risco0} has no real solutions and stable circular orbits exist all the way from spatial infinity to the singularity
(see \cite{TMD}).

\begin{figure*}[t]
\begin{minipage}{0.48\linewidth}
\center{\includegraphics[width=0.97\linewidth]{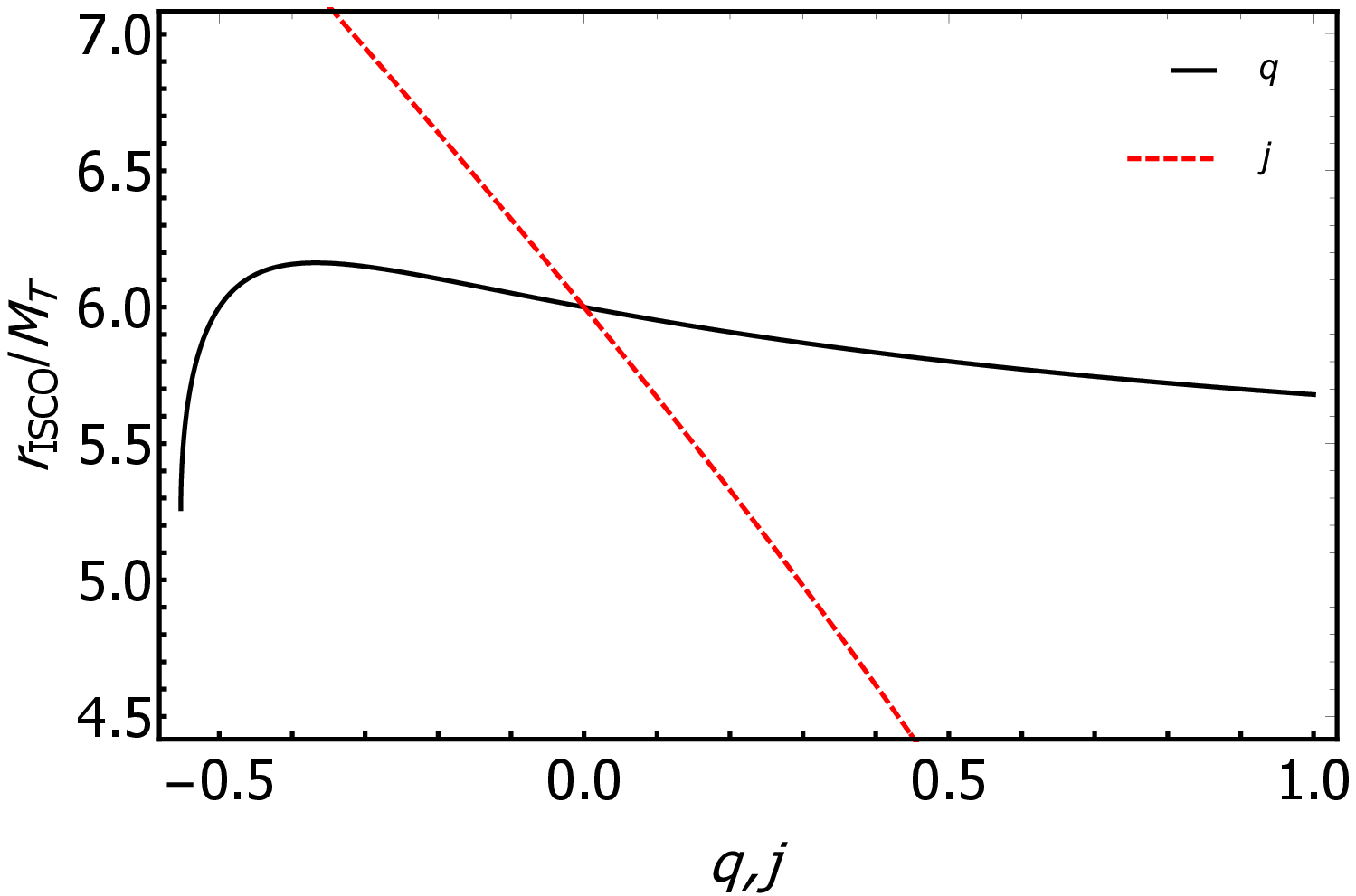}\\ } 
\end{minipage}
\hfill 
\begin{minipage}{0.51\linewidth}
\center{\includegraphics[width=0.97\linewidth]{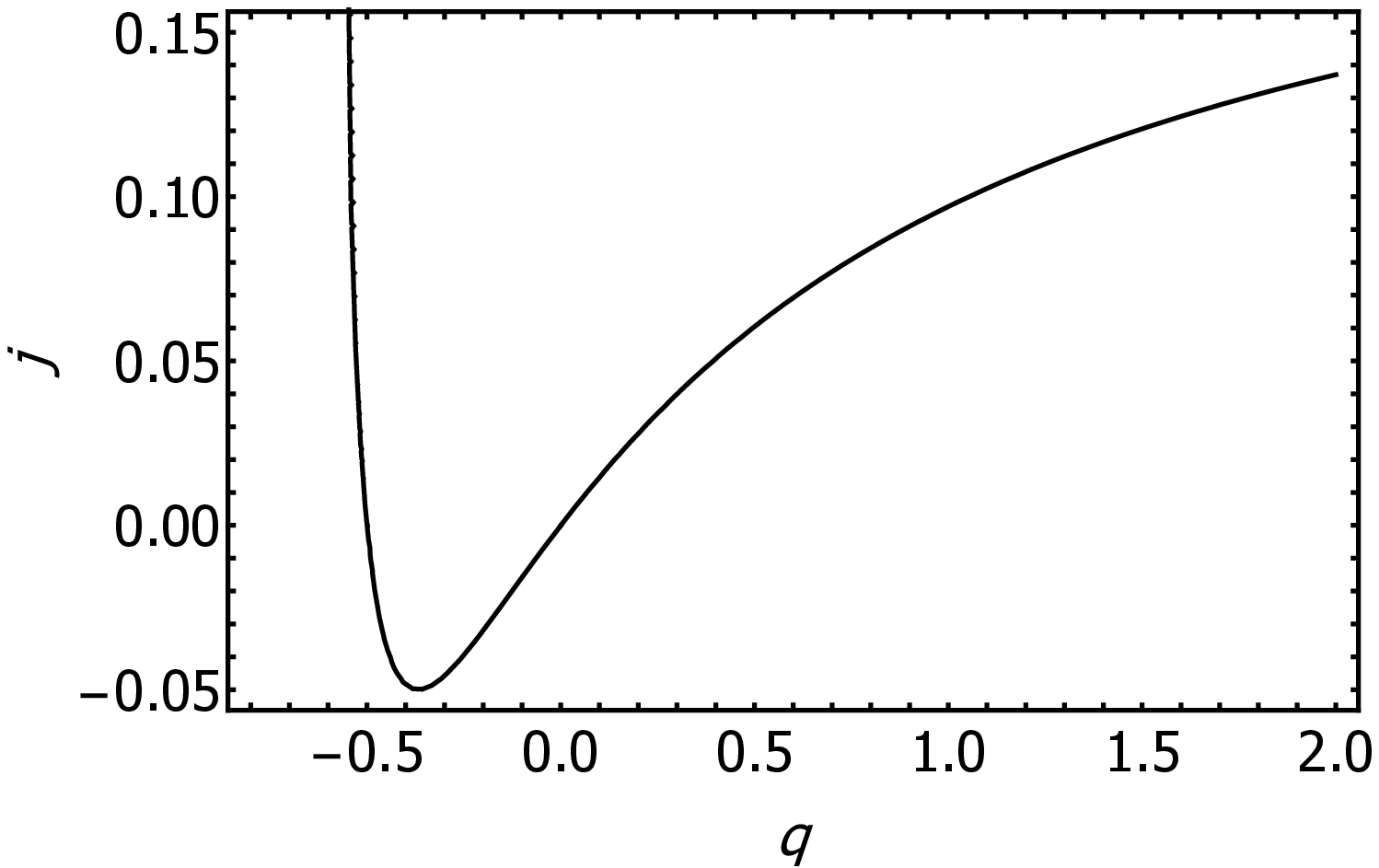}\\ }
\end{minipage}
\caption{Left panel: The ISCO radii for the $q$-metric in units of total mass $M_T$ as a function of the quadrupole parameter $q$ as compared to the ISCO for the Kerr space-time as a function of the dimensionless angular momentum $j$. Right panel: Degeneracy for the value of the ISCO between the $q$-metric and the Kerr space-time. Namely for each allowed value of $q$ there is a corresponding value of $j$ for which the two sources with the same active mass $M_T$ produce the same ISCO.}
\label{fig:riscoqj}
\end{figure*}

Notice that in order to compare the ISCO with that of the Kerr space-time we must consider the two geometries with the source of the same active gravitational mass. The total mass of the $q$-metric may be evaluated via Komar's integral and it is given by $M_T=m(1+q)$ and therefore the ISCO radius as a function of $q$ in units of total mass $M_T$ is given by
\be 
\frac{r_{ISCO}}{M_T}=3+\frac{1}{1+q}+\sqrt{5-\frac{1}{(1+q)^2}},
\ee 
and it is shown in the left panel of Fig.~\ref{fig:riscoqj}. The right panel of Fig.~\ref{fig:riscoqj} shows the degeneracy between the ISCO radius in the $q$-metric as a function of $q$ and that of the Kerr metric as a function of the dimensionless angular momentum $j$.

\section{Particle motion in the Kerr metric}\label{sez2b}

The Kerr space-time  \cite{1963PhRvL..11..237K}
is a vacuum geometry describing a rotating uncharged axially-symmetric black hole. The line element for the Kerr metric in Boyer-Lindquist coordinates is
\begin{eqnarray}
ds^2&=&\left(1-\frac{2Mr}{\Sigma}\right)dt^{2}-\frac{\Sigma}{\Delta}dr^{2}+\frac{4Mra}{\Sigma}\sin^{2}\theta d\phi dt\nonumber\\
&-&\Sigma d\theta^{2}-\left(r^{2}+a^{2}+\frac{2Mra^{2}}{\Sigma}\sin^2\theta\right)d\phi^{2},\
\label{eq:metricKerr}
\end{eqnarray}
where we have set $\Sigma=r^{2}+a^{2}\cos^{2}\theta$ and $\Delta=r^{2}-2Mr+a^{2}$. The total gravitational mass of the source is given by $M_T=M$ and its dimensionless angular momentum is $j=a/M$. Therefore, also the Kerr metric is fully characterized by two parameters only and Schwarzschild is obtained for vanishing angular momentum, i.e. $a=0$.

The Kerr space-time is the paradigm for describing the exterior of astrophysical black hole candidates, however as of now no precise measurement of the actual geometry of such objects is available. For this reason it is important to keep an open mind and allow for the possibility that such objects may be described by more exotic solutions such as the $q$-metric. Therefore it is worth considering what kinds of observations may allow to distinguish the two sources. As mentioned earlier, most of the existing black hole candidates are observed through the spectra of their accretion disks and therefore in the following we shall review circular orbits and ISCO for the Kerr solution with the aim of comparing the results with the corresponding ones in the $q$-metric.

\subsection{Circular orbits in the Kerr metric}

The specific angular velocity, angular momentum and energy of the particle moving on a circular orbit around a Kerr black hole are derived, respectively as

\bea 
 \label{eq:omegaKerr}
\Omega^{2}&=& \frac{M}{r^{3}\pm 2a r^{2}\sqrt{M/r} +a^{2}M}, \\
 \label{eq:energyKerr}
E^{2}&=&\frac{\left(\sqrt{r}\left(r-2M \right)\pm a \sqrt{M}\right)^{2}}{r^{2}\left(r\pm 2a\sqrt{M/r}-3M\right)}, \\
 \label{eq:angmomKerr}
L^{2}&=& \frac{M\left(r^{2}\mp 2a\sqrt{M/r}+a^{2}\right)^{2}}{r^{2}\left(r\pm 2a\sqrt{M/r}-3M\right)},
\eea
where the $+$ and $-$ signs correspond to co-rotating and
counter-rotating particles with respect to the direction of rotation of the black hole. 

Analogously to the $q$-metric, we write the radius of the ISCO for the Kerr metric as \cite{1972ApJ...178..347B}
\begin{equation}
 \label{eq:riscoKerr}
\frac{r_{ISCO}^{\pm}}{M_T}=\left(3+Z_2\mp \sqrt{(3-Z_1)(3+Z_1+2Z_2)}\right),   
\end{equation}
where we have defined

\begin{subequations}
\begin{align}
Z_1&\equiv 1+\left(1-j^2\right)^{\frac{1}{3}}\left(\left(1+j\right)^{\frac{1}{3}}+\left(1-j\right)^{\frac{1}{3}}\right),\label{eq:Z1}\\ 
Z_2&\equiv \left(3 j^2+Z^2_1\right)^{\frac{1}{2}}.\label{eq:Z2} 
\end{align}
\end{subequations}

Having obtained circular orbits and the ISCO in both space-times (see Fig.~\ref{fig:riscoqj} for details) we are now able to consider the true observables that may be obtained from astrophysical black hole candidates, namely the spectra of light emitted from the accretion disks.

\begin{figure*}[ht]
\begin{minipage}{0.49\linewidth}
\center{\includegraphics[width=0.97\linewidth]{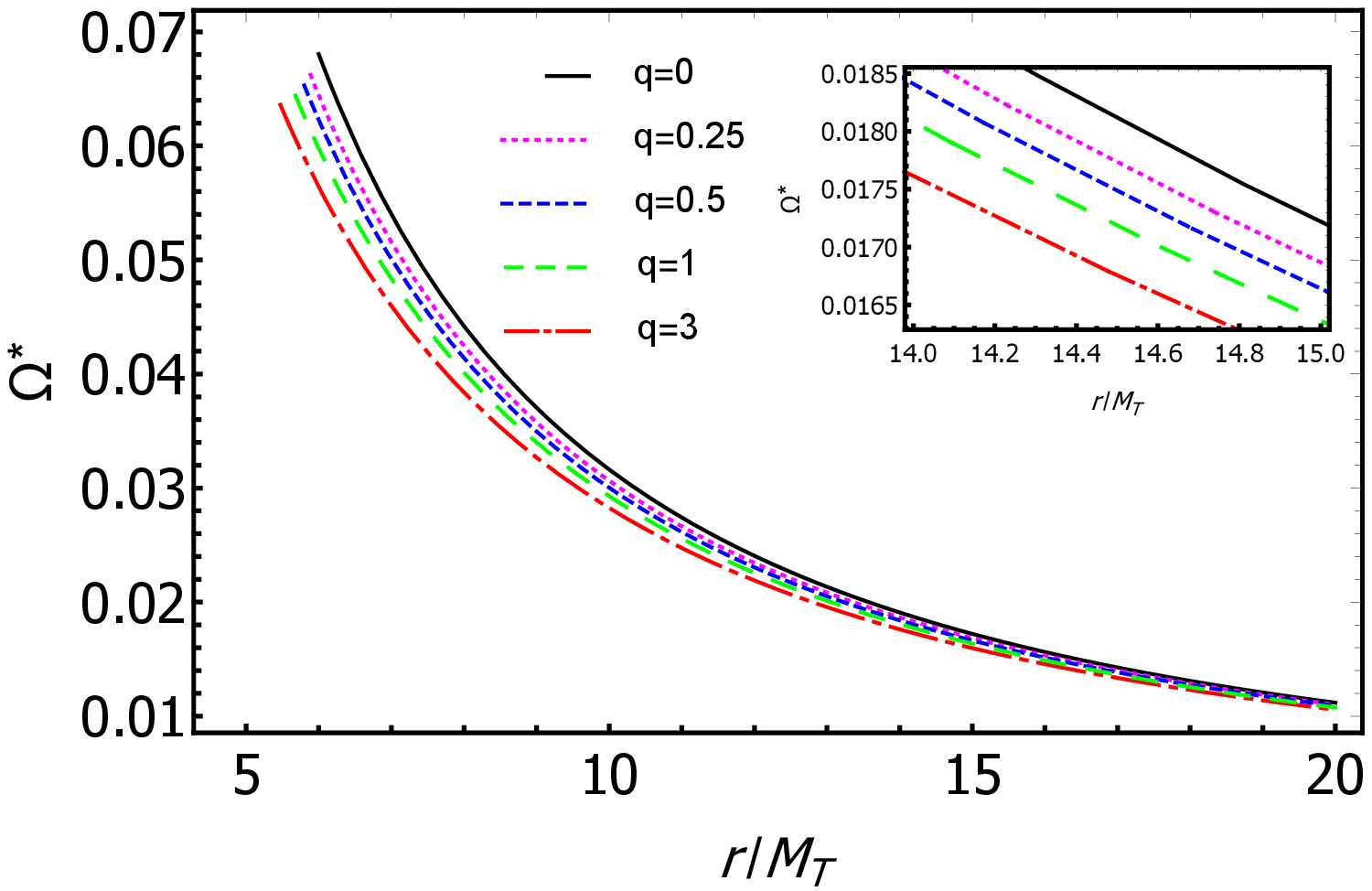}\\ }
\end{minipage}
\hfill 
\begin{minipage}{0.50\linewidth}
\center{\includegraphics[width=0.97\linewidth]{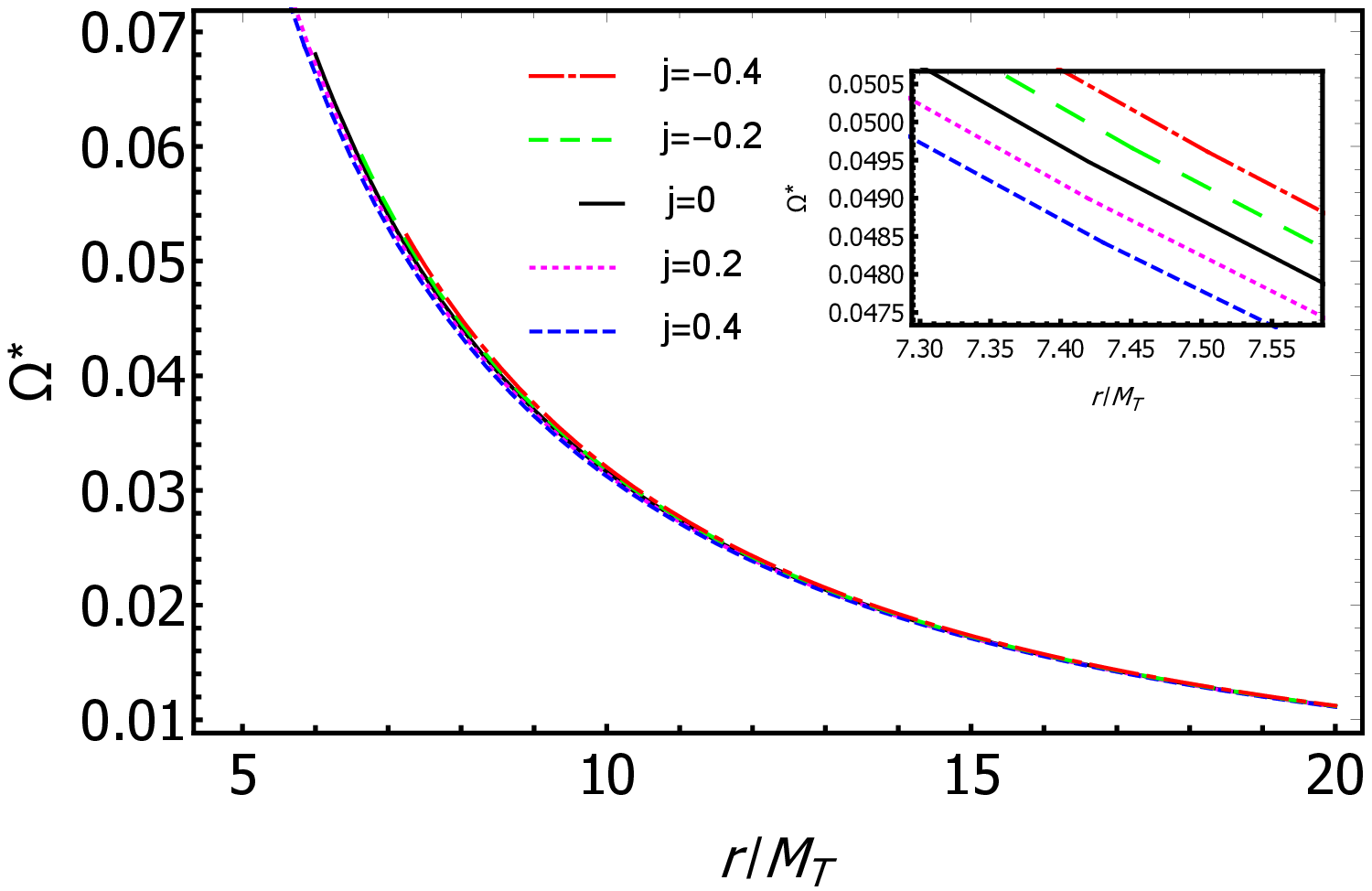}\\ }
\end{minipage}
\caption{Color online. Left panel: Angular velocity of test particles versus radial distance $r$ normalized in units of total mass $M_T$ in the oblate $q$-metric. Right panel: Angular velocity of test particles versus radial distance $r$ normalized in units of total mass $M_T$ in the Kerr space-time.}
\label{fig:omega}
\end{figure*}

\section{Spectra of thin accretion disks}\label{sez3}

In order to study the luminosity and spectrum of the accretion disk in the $q$-metric we follow the model proposed by Novikov-Thorne and Page-Thorne in \cite{novikov1973, page1974}.

Accordingly, the radiative flux $\mathcal{F}$ (i.e. the energy radiated per unit area per unit time) emitted by the accretion disk is given by 

\begin{equation}
 \label{eq:flux}
\mathcal{F}(r)=-\frac{\dot{{\rm m}}}{4\pi \sqrt{g}} \frac{\Omega_{,r}}{\left(E-\Omega L\right)^2 }\int^r_{r_{ISCO}} \left(E-\Omega L\right) L_{,\tilde{r}}d\tilde{r},
\end{equation}
where $\dot{{\rm m}}$ is the mass accretion rate of the disk, which is assumed to be constant and $g$ is the determinant of the metric tensor of the three-dimensional sub-space ($t,r,\varphi$), i.e. $\sqrt{g}=\sqrt{g_{tt}g_{rr}g_{\varphi\varphi}}$.

Furthermore, it must be noticed that $\mathcal{F}$ is not directly observable since it is a local quantity that is measured in the rest frame of the disk. Therefore a more interesting quantity from the observational perspective is the differential luminosity (i.e. the energy per unit time that reaches an observer at infinity) $\mathcal{L}_{\infty}$ which can be estimated from the flux $\mathcal{F}$ via the following relation \cite{novikov1973, page1974}
\begin{equation}
 \label{eq:difflum}
\frac{d\mathcal{L}_{\infty}}{d\ln{r}}=4\pi r \sqrt{g}E \mathcal{F}(r).
\end{equation}

Both of these quantities describe the amount of radiation emitted by the disk at a given radius $r$. However, what is measured in practice is the spectrum of the light emitted as a function of the frequency. Therefore, another characteristic quantity of the accretion disk that is worth considering is the spectral luminosity distribution observed at infinity $\mathcal{L}_{\nu,\infty}$. Under the assumption of black body emission from the accretion disk $\mathcal{L}_{\nu,\infty}$ is given by \cite{2020MNRAS.496.1115B}
\begin{equation}
 \label{eq:speclum}
\nu \mathcal{L}_{\nu,\infty}=\frac{60}{\pi^3}\int^{\infty}_{r_{ISCO}}\frac{\sqrt{g }E}{M_T^2}\frac{(u^t y)^4}{\exp\left[u^t y/\mathcal{F}^{*1/4}\right]-1}dr,
\end{equation}
where $u^t$ is the co-variant time component of the four velocity, defined by
\begin{equation}
 \label{eq:sample7}
u^t(r)=\frac{1}{\sqrt{g_{tt}+2\Omega g_{t\varphi}+\Omega^2 g_{\varphi \varphi}}},
\end{equation}
and $y=h\nu/kT_*$, $h$ is the Planck constant, $\nu$ is the frequency of the emitted radiation, $k$ is the Boltzmann constant, $T_*$ is the characteristic temperature as defined from Stefan-Boltzmann law as
$
\sigma T_*=\dot{{\rm m}}/4\pi M_T^2
$,
with $\sigma$ being Stefan-Boltzmann constant.
Notice that to keep the argument of the exponential dimensionless we have normalized the flux with respect to total mass $M_T$ and defined $\mathcal{F}^*(r)=M_T^2\mathcal{F}(r)$.

Finally, another quantity of interest related to the radiation emitted by the accretion disk is the net luminosity that reaches observers at infinity $\mathcal{L}_{\infty}$, which is given by the integral of equation \eqref{eq:difflum}. This quantity represents the amount of rest mass energy of the accreting matter that is converted into radiation and can also be expressed as $\mathcal{L}_{\infty}=(1-E_{ISCO})\dot{{\rm m}}$, or, in other words, the efficiency of the source in converting infalling mass into emitted radiation. For Schwarzschild, taking unit mass accretion rate it is known that $(1-E_{ISCO})=0.0572$, meaning that an accretion disk around a Schwarzschild black hole converts matter into radiation with an efficiency of $5.72\%$,

\noindent 
\begin{figure*}[ht]
\begin{minipage}{0.49\linewidth}
\center{\includegraphics[width=0.97\linewidth]{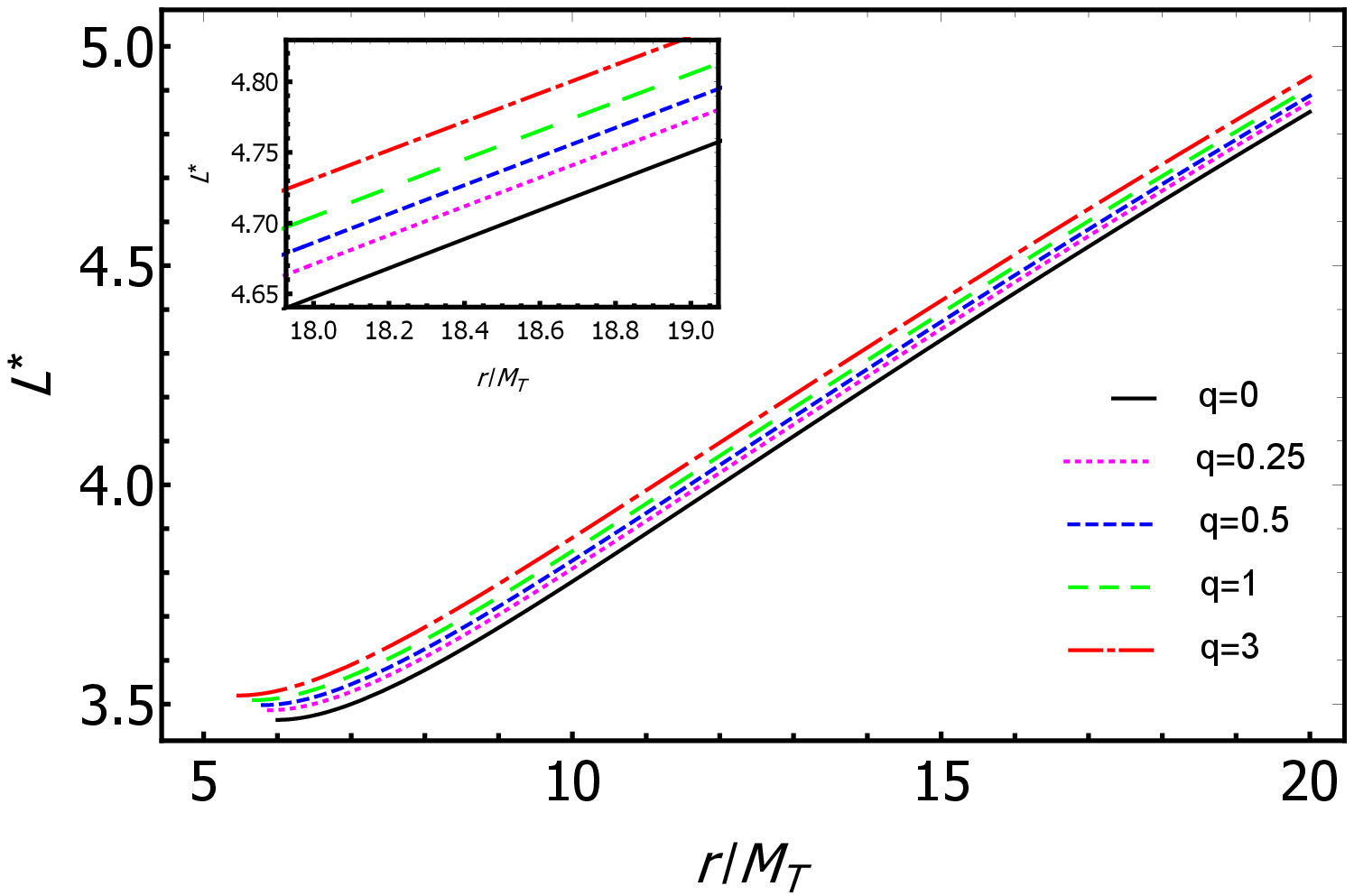}\\ } 
\end{minipage}
\hfill 
\begin{minipage}{0.50\linewidth}
\center{\includegraphics[width=0.97\linewidth]{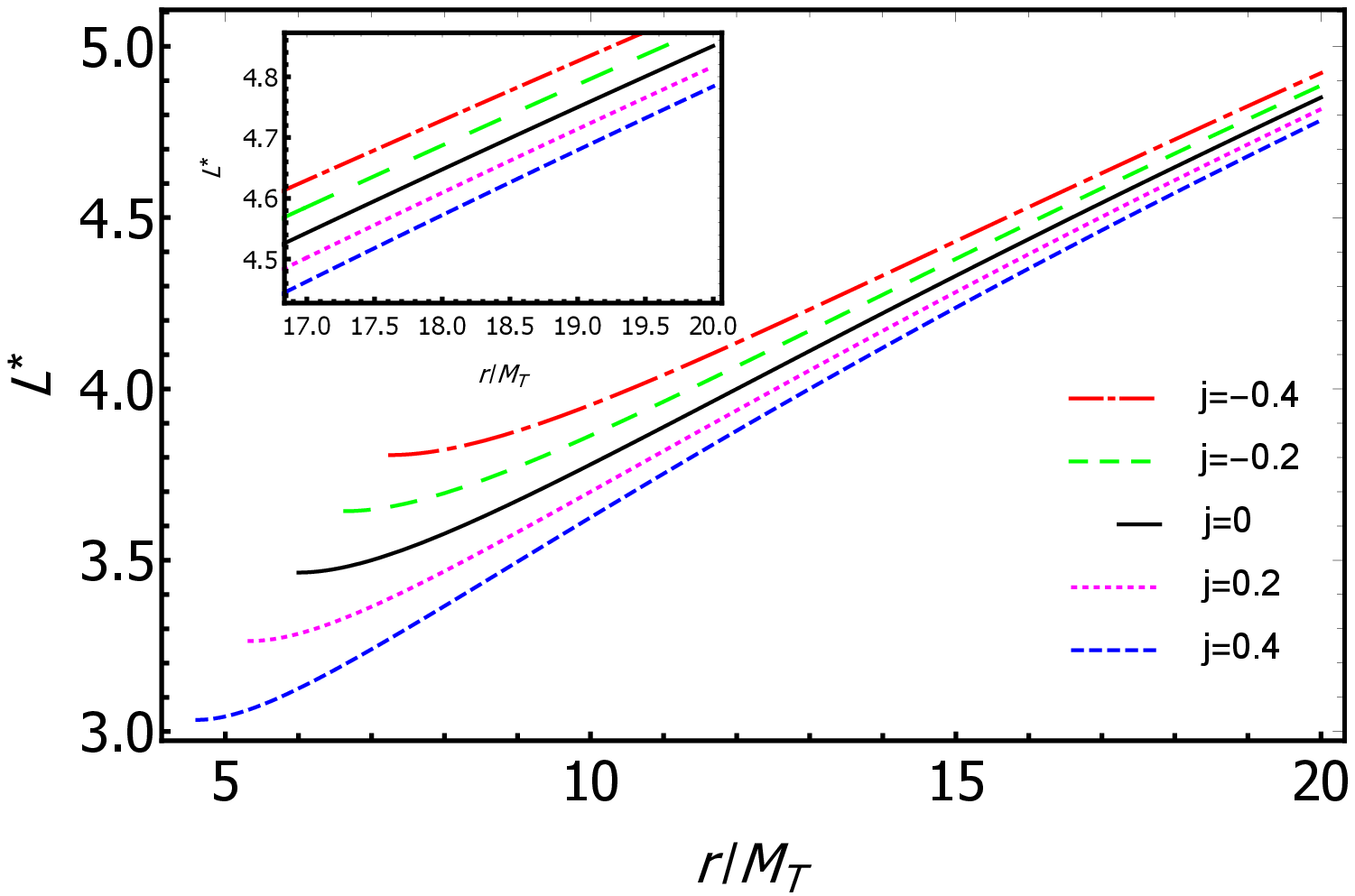}\\ }
\end{minipage}
\caption{Color online. Angular momentum $L^*$ of test particles versus radial distance $r$ normalized in units of total mass $M_T$. Left panel: $L^*$ in the $q$-metric for oblate sources ($q>0$). Right panel: in the Kerr space-time.}
\label{fig:angmom}
\end{figure*}
\begin{figure*}[ht]
\begin{minipage}{0.49\linewidth}
\center{\includegraphics[width=0.97\linewidth]{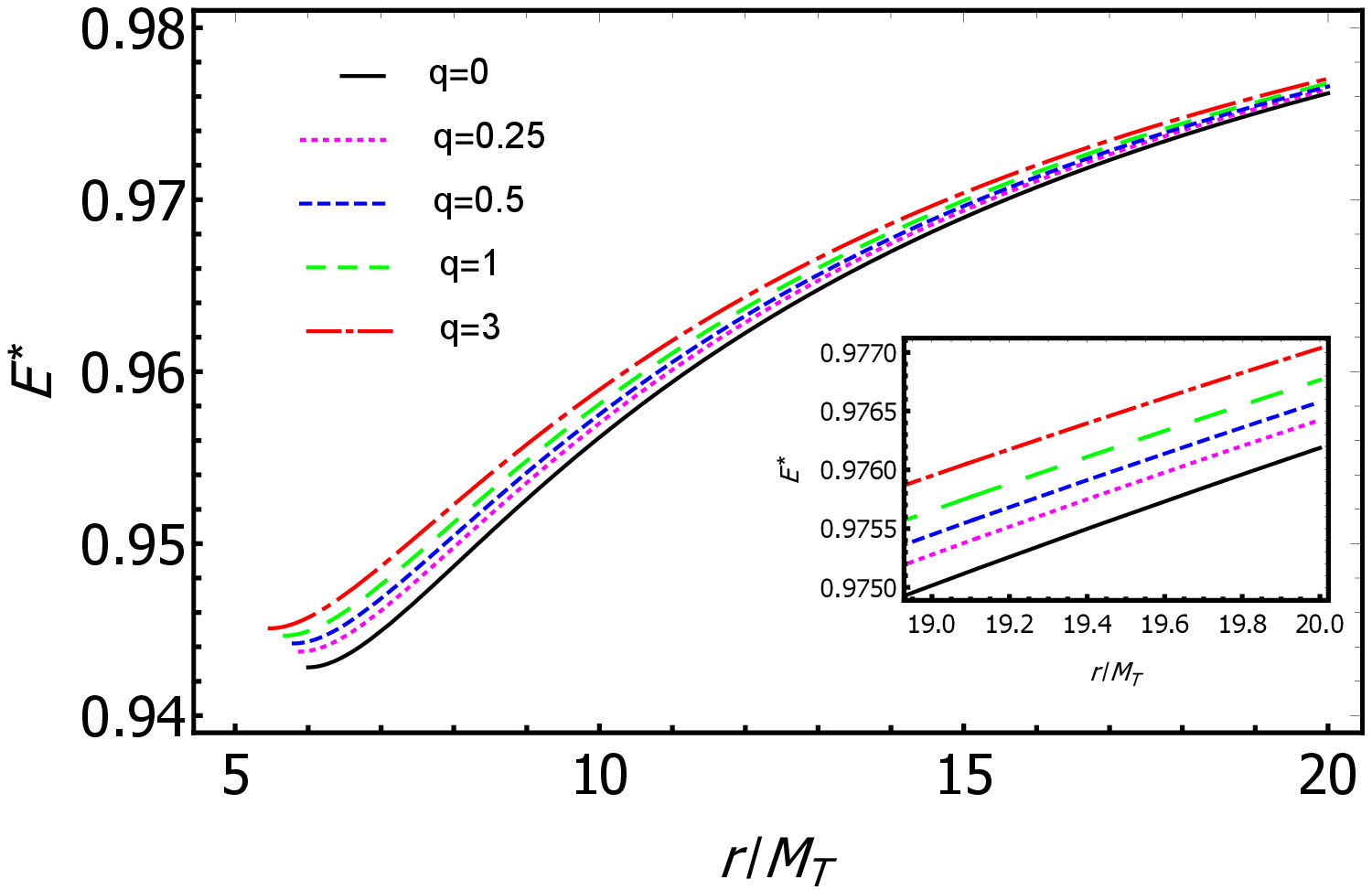}\\ } 
\end{minipage}
\hfill 
\begin{minipage}{0.50\linewidth}
\center{\includegraphics[width=0.97\linewidth]{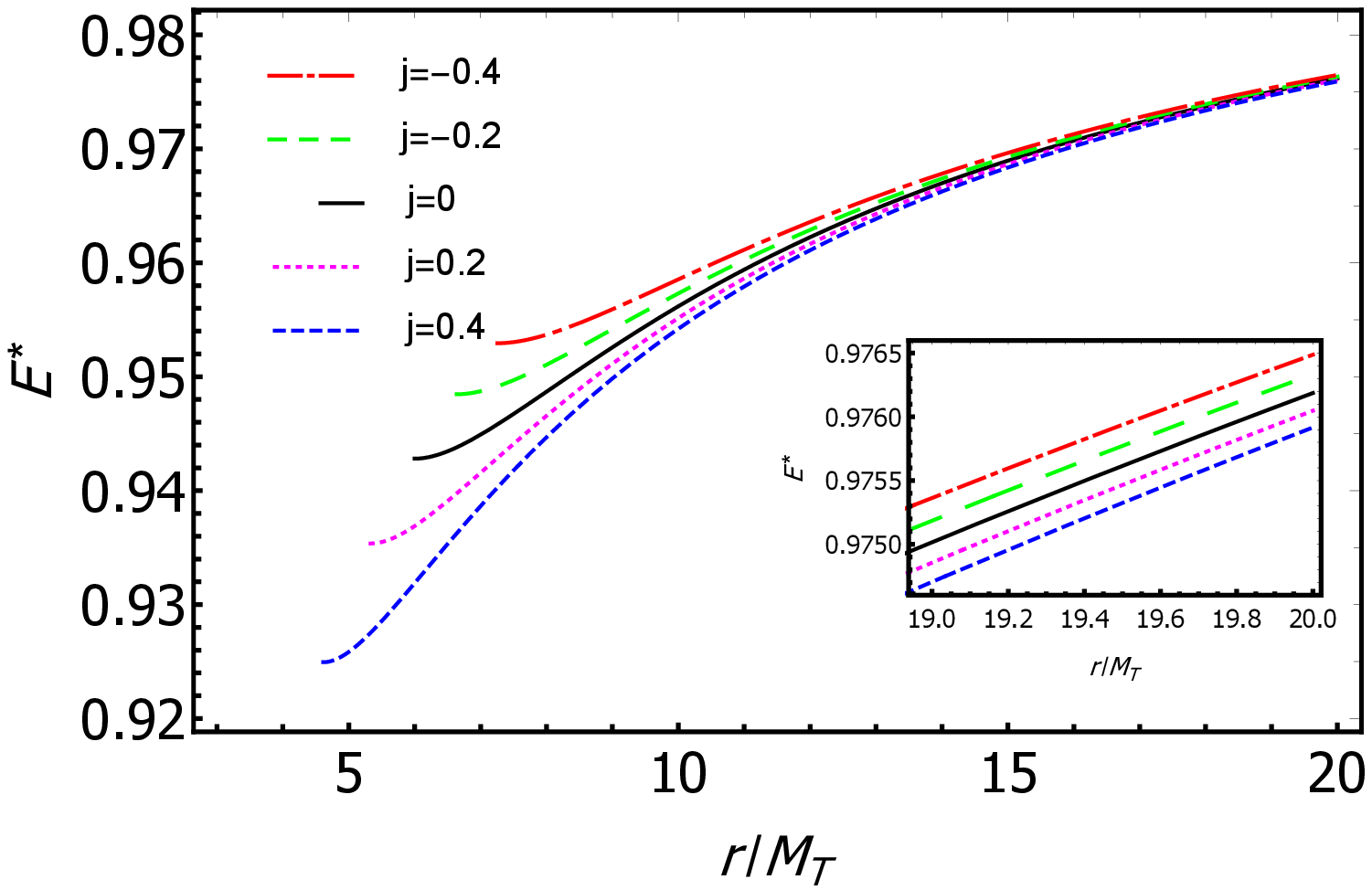}\\ }
\end{minipage}
\caption{Color online. Energy $E^*$ of test particles versus radial distance $r$ normalized in units of total mass $M_T$. Left panel: $E^*$ in the oblate $q$-metric. Right panel: $E^*$ in the Kerr space-time.}
\label{fig:energy}
\end{figure*}
\begin{figure*}[ht]
\begin{minipage}{0.49\linewidth}
\center{\includegraphics[width=0.97\linewidth]{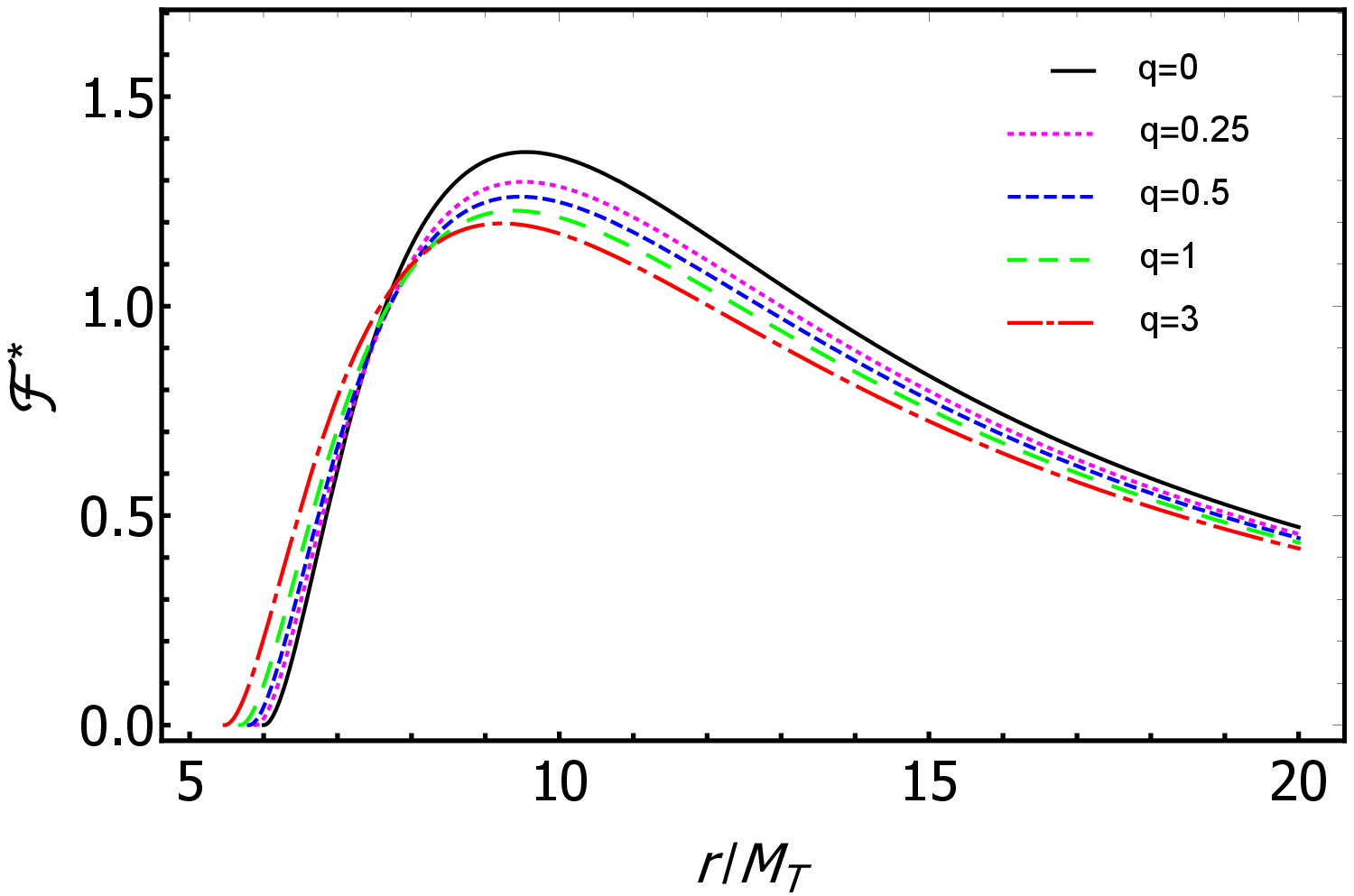}\\ } 
\end{minipage}
\hfill 
\begin{minipage}{0.50\linewidth}
\center{\includegraphics[width=0.97\linewidth]{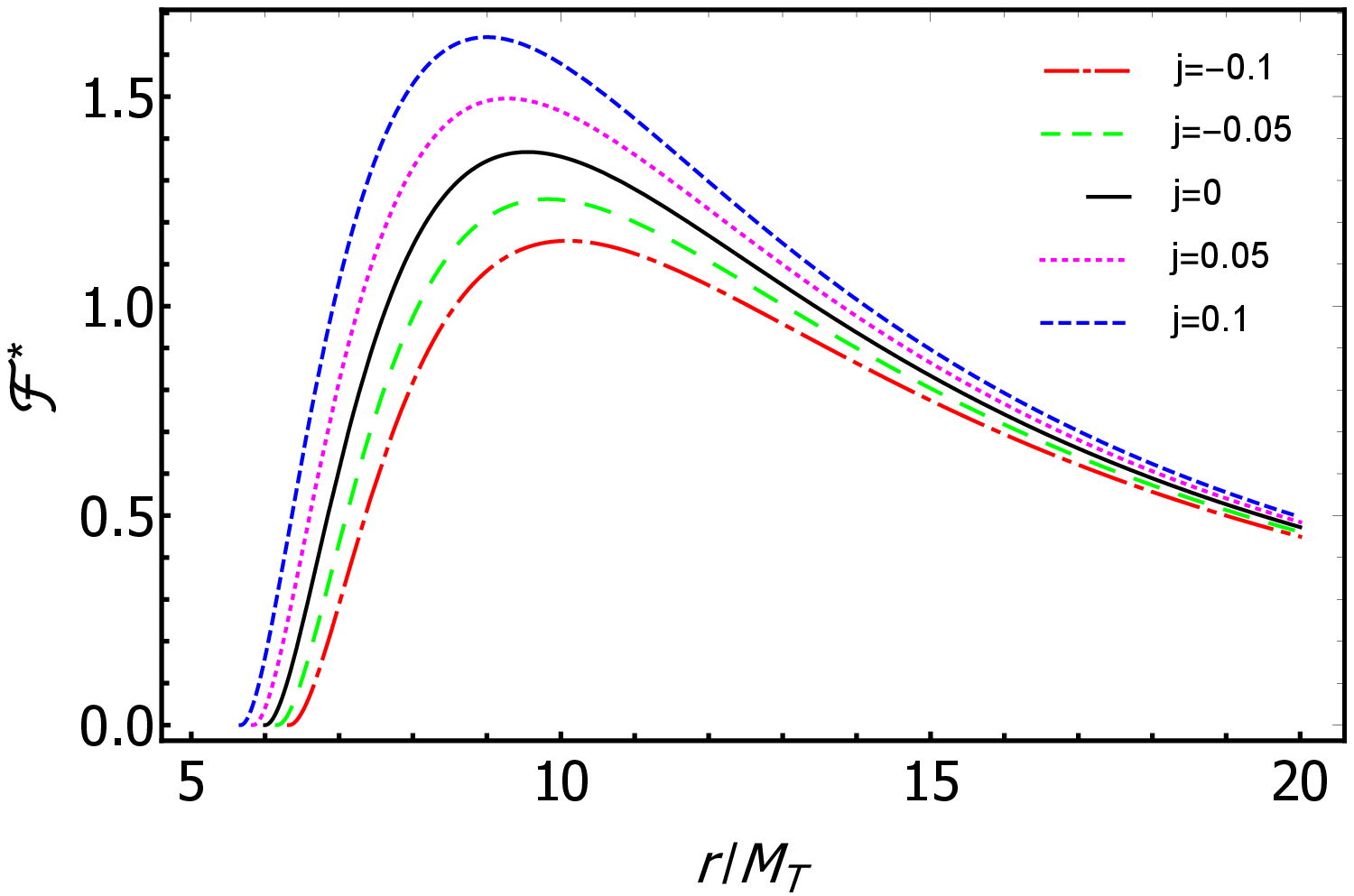}\\ }
\end{minipage}
\caption{Color online. Radiative flux $\mathcal{F}^*$ multiplied by $10^5$ of the accretion disk versus radial distance $r$ normalized in units of total mass $M_T$. Left panel: $\mathcal{F}^*$ in the oblate $q$-metric. Right panel: $\mathcal{F}^*$ in the Kerr space-time. Notice that for the Kerr metric the change in $j$ causes the flux to increase or decrease everywhere, while for the oblate $q$-metric the flux increases at small radii and decreases at large radii with respect to Schwarzschild.}
\label{fig:flux}
\end{figure*}

\subsection{Numerical results}\label{sez4}

In the limit of negligible rotation, the space-time in the exterior of a compact object may be approximated as static. At the same time the deformation effects on the source would introduce some oblateness. Therefore, in the following we have considered only oblate sources (i.e. $q>0$) as they are more physically relevant in describing the shape of a rotating object. Also, it must be mentioned that, in order to carry out the numerical analysis, it is useful to introduce dimensionless quantities defined as $\Omega^*(r)=M_T\Omega(r)$, $L^*(r)=L(r)/M_T$ and $E^*(r)=E(r)$.

In Fig.~\ref{fig:omega} (left panel) we show the orbital angular velocity $\Omega^*(r)$ of test particles as a function of the normalized radial coordinate $r/M_T$ in the $q$-metric. It can be seen that the inclusion of the quadrupole moment for oblate sources noticeably decreases $\Omega^*(r)$ with respect to the Schwarzschild ($q=0$) case. For comparison we plot  $\Omega^*(r)$ in the Kerr space-time \cite{1963PhRvL..11..237K} on the right panel of Fig.~\ref{fig:omega}. Notice that in the plots of particles on circular orbits in the Kerr metric we have used the dimensionless angular momentum or spin parameter defined by $j=a/M_T$, where $a$ is the rotation Kerr parameter and $M_T$ is the total mass of the Kerr metric as mentioned above. Obviously, for vanishing $j$ one recovers the Schwarzschild metric.

In Fig.~\ref{fig:angmom} (left panel) we constructed the dimensionless orbital angular momentum $L^*(r)$ of test particles as a function of the normalized radial coordinate in the $q$-metric. We see that for different values of $q$ angular momentum is always larger than the Schwarzschild $q=0$ case and a similar behavior is observed in the Kerr metric for counter-rotating particles, i.e $j<0$ (right panel).

In Fig.~\ref{fig:energy} (left panel) we constructed the energy per unit mass $E^*$ of test particles as a function of the normalized radial coordinate in the $q$-metric. Here $E^*$ is always larger than in the Schwarzschild case and again the behavior is mimicked in the Kerr space-time by counter-rotating particles (right panel). 

It is easily noticed that the motion of counter-rotating ($j<0$) particles on circular orbits in the Kerr metric is mimicked by the effects of the oblateness ($q>0$) of the $q$-metric. Similarly one could see that co-rotating particles are mimicked by prolate sources of the $q$-metric~\cite{TMD}.

Due to the differences in $\Omega^*$, $L^*$ and $E^*$ between the oblate sources of the $q$-metric and the Kerr metric with co-rotating disk, we expect that the radiative flux emitting from the accretion disks will be different. However, we expect accretion disks around the oblate $q$-metric to mimic the counter-rotating disk in the Kerr space-time. 
In order to check this, we plotted the radiative flux versus normalized radial coordinate in Fig.~\ref{fig:flux} (left panel) in the $q$-metric. Some noticeable differences appear in the vicinity of the ISCO as $\mathcal{F}^*$ in the $q$-metric is larger than than the Schwarzschild case for small radii due to the fact that the ISCO radius is smaller, while for the Kerr space-time we have smaller flux for counter-rotating disk at all radii (right panel). 
 
On the other hand, as expected, for the co-rotating disk in the Kerr metric (right panel) the flux is everywhere larger than in the Schwarzschild metric. 
This suggests that accretion flux may be able to distinguish disks surrounding static oblate sources from co-rotating disks around a Kerr black hole. 

\begin{figure*}[ht]
\begin{minipage}{0.49\linewidth}
\center{\includegraphics[width=0.97\linewidth]{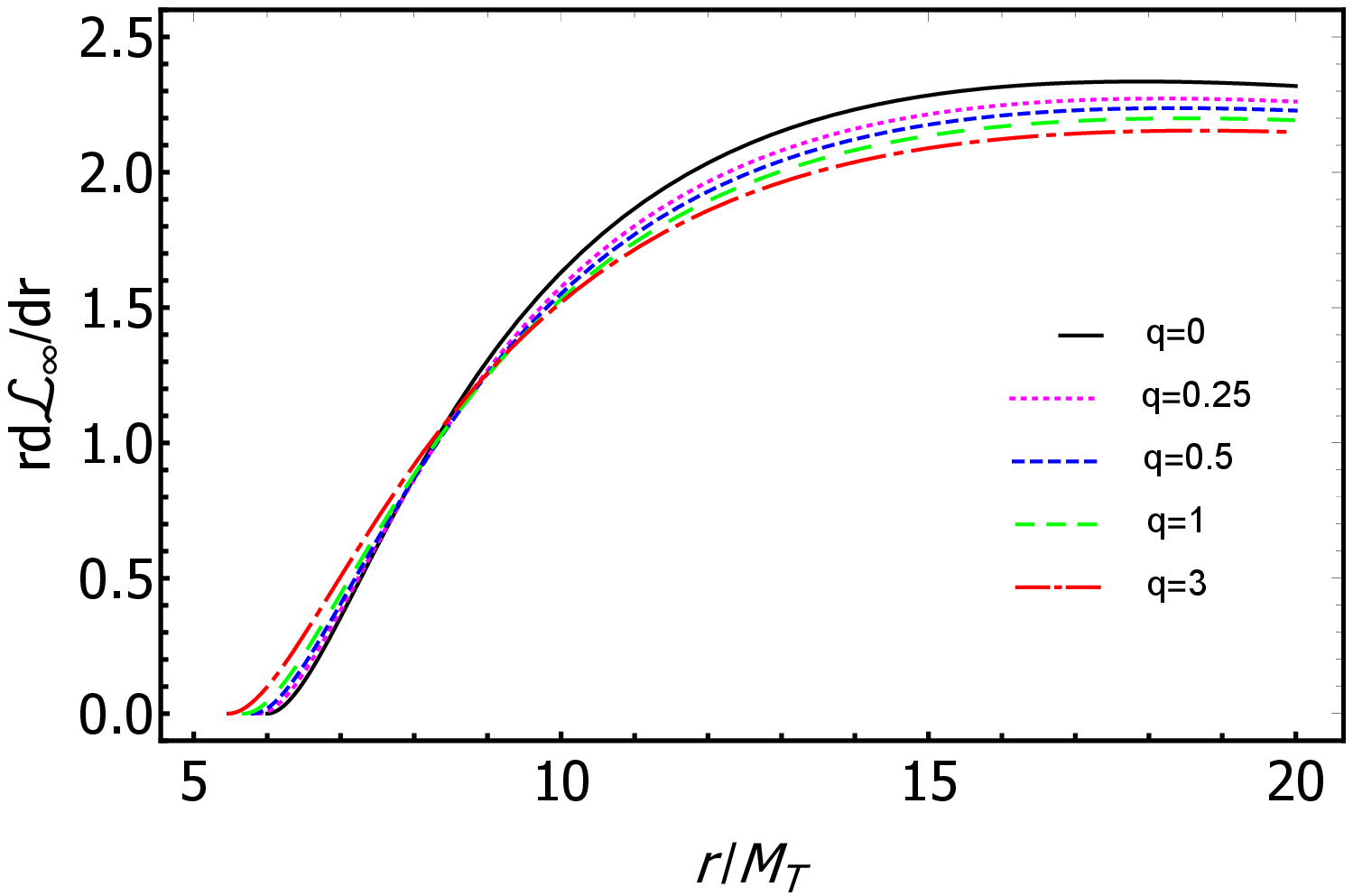}\\ } 
\end{minipage}
\hfill 
\begin{minipage}{0.50\linewidth}
\center{\includegraphics[width=0.97\linewidth]{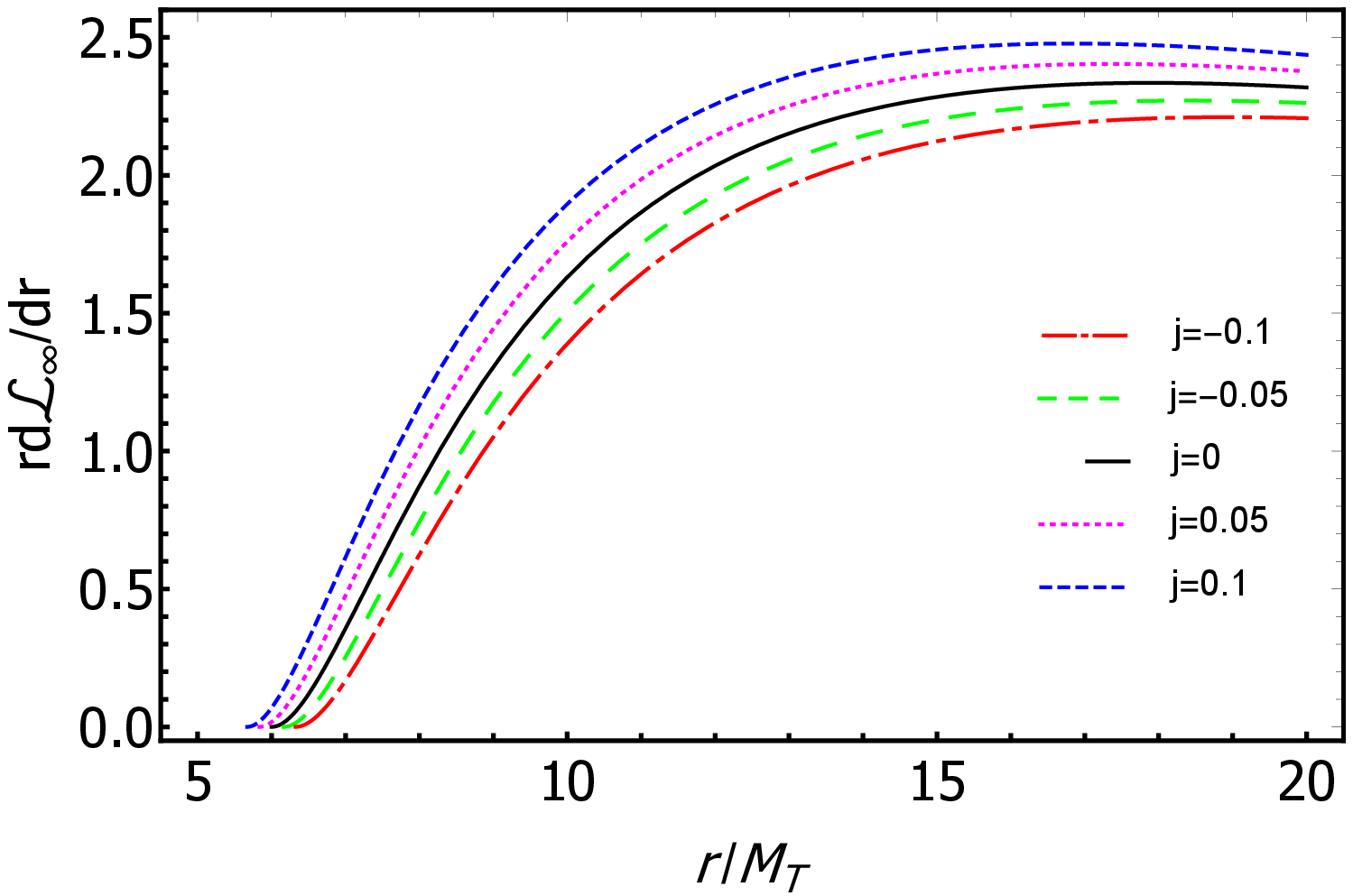}\\ }
\end{minipage}
\caption{Color online. Differential luminosity multiplied by $10^2$ of the accretion disk versus radial distance $r$ normalized in units of total mass $M_T$. Left panel: Differential luminosity in the oblate $q$-metric. Right panel: Differential luminosity in the Kerr space-time.}
\label{fig:difflum}
\end{figure*}
\begin{figure*}[ht]
\begin{minipage}{0.49\linewidth}
\center{\includegraphics[width=0.97\linewidth]{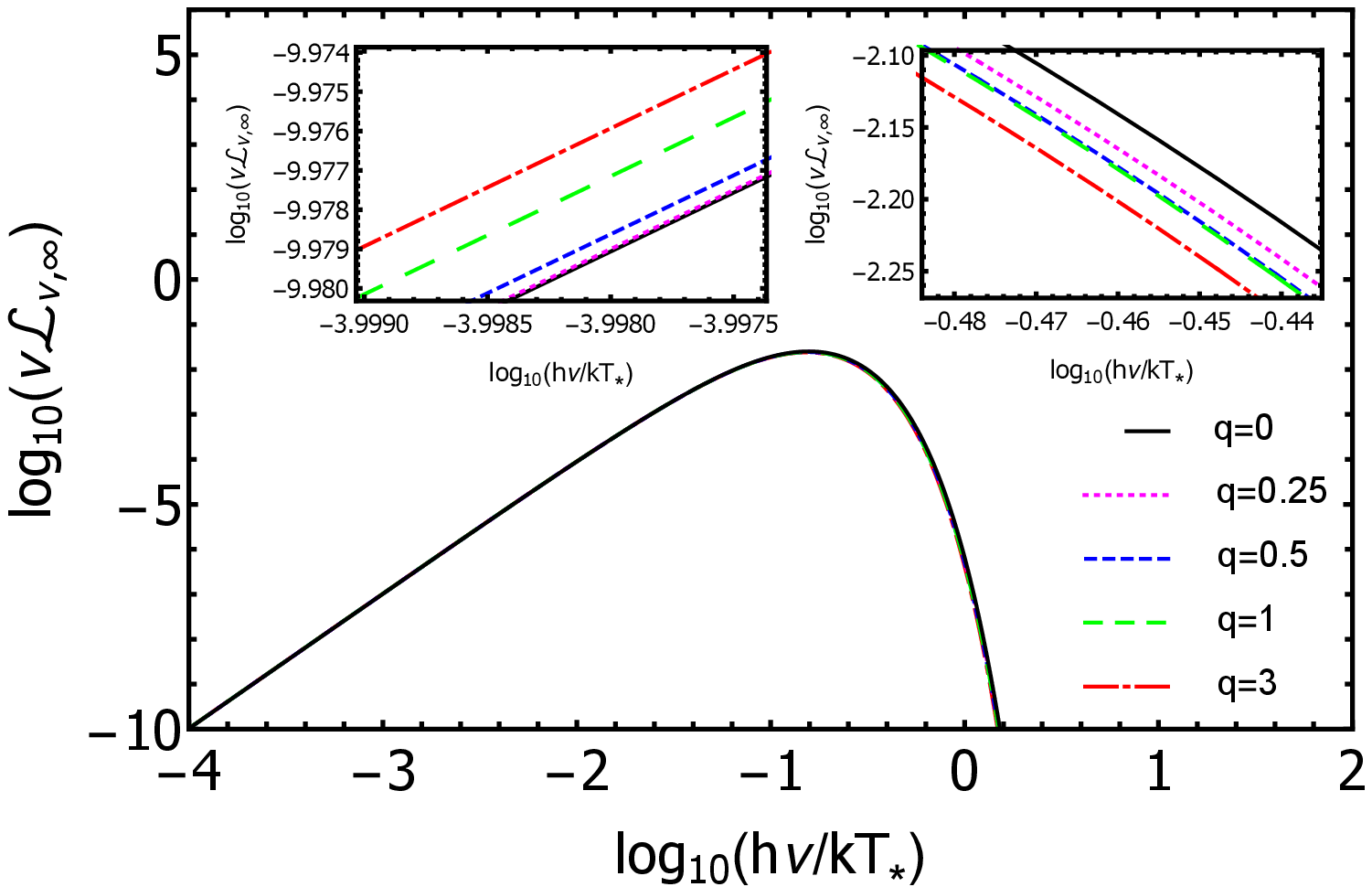}\\ } 
\end{minipage}
\hfill 
\begin{minipage}{0.50\linewidth}
\center{\includegraphics[width=0.97\linewidth]{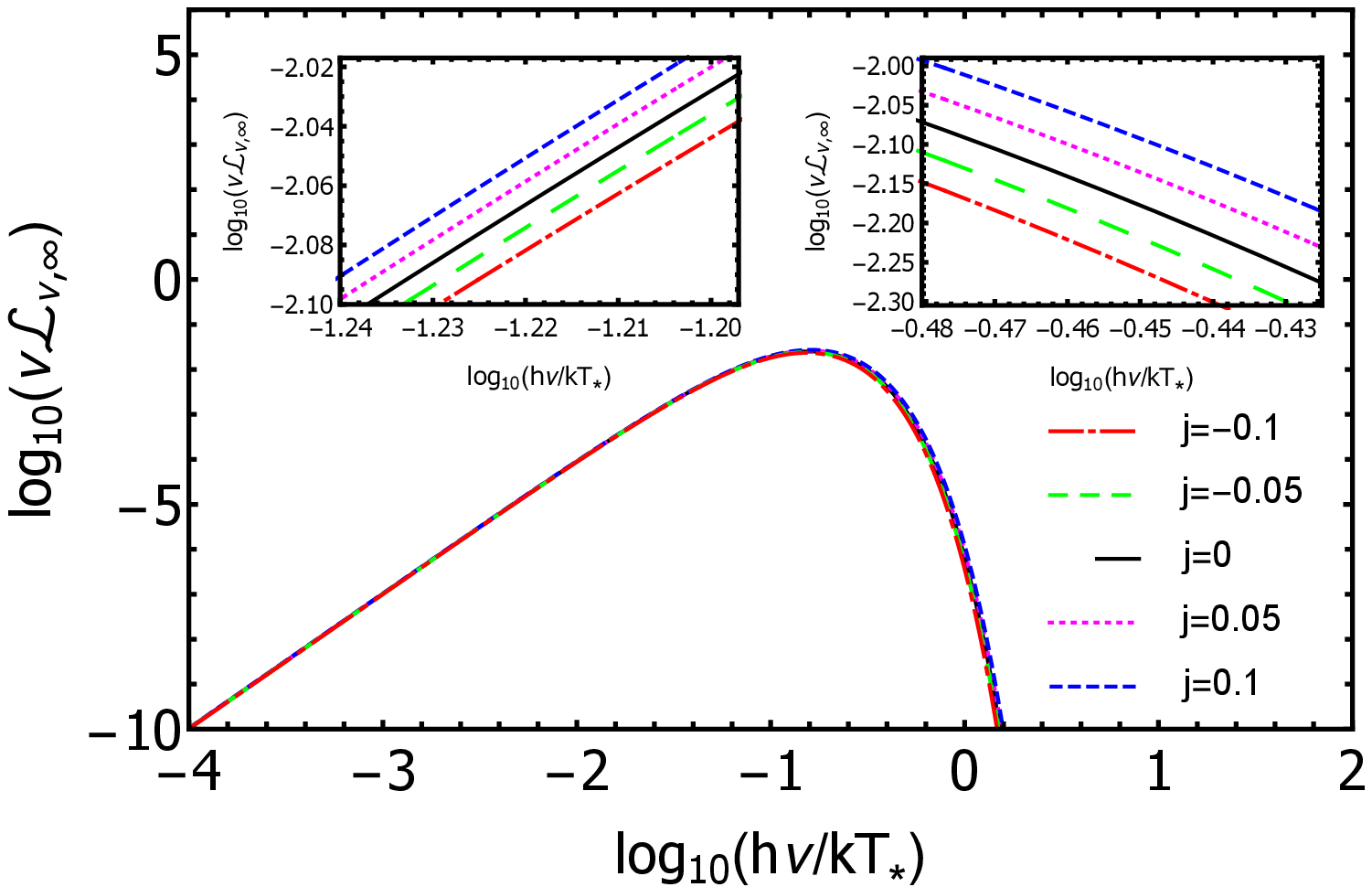}\\ }
\end{minipage}
\caption{Color online. Spectral luminosity versus frequency of the emitted radiation for blackbody emission of the accretion disk. Left panel: Spectral luminosity in the oblate $q$-metric. Right panel: Spectral luminosity in the Kerr space-time.}
\label{fig:speclum}
\end{figure*}

In Fig.~\ref{fig:difflum} (left panel) we show the differential luminosity versus normalized radial coordinate in the oblate $q$-metric and compare it with the corresponding quantity in the Kerr metric (right panel). Since the differential luminosity is defined via the flux we see that the behavior observed in Fig.~\ref{fig:flux} translates into the differential luminosity. More precisely, the differential luminosity for the oblate $q$-metric is larger than that of Schwarzschild only for small radii and becomes smaller at larger distances from the source, while for the Kerr case the differential luminosity is everywhere smaller (larger) than Schwarzschild's for the counter-rotating (co-rotating) case.

In Fig.~\ref{fig:speclum} (left panel) we plotted the spectral luminosity $\mathcal{L}_{\nu,\infty}$ as defined by Eq.~\eqref{eq:speclum} as a function of the frequency of radiation emitted by the accretion disk in the $q$-metric. For small frequencies the spectral luminosity is larger than in the Schwarzschild case and vice versa. In the Kerr metric (right panel) the spectral luminosity is always larger (smaller) than in the Schwarzschild metric for co-rotating (counter-rotating) disks.

\begin{figure*}[ht]
\centering
\begin{tabular}{lr}
\includegraphics[width=0.5\linewidth]{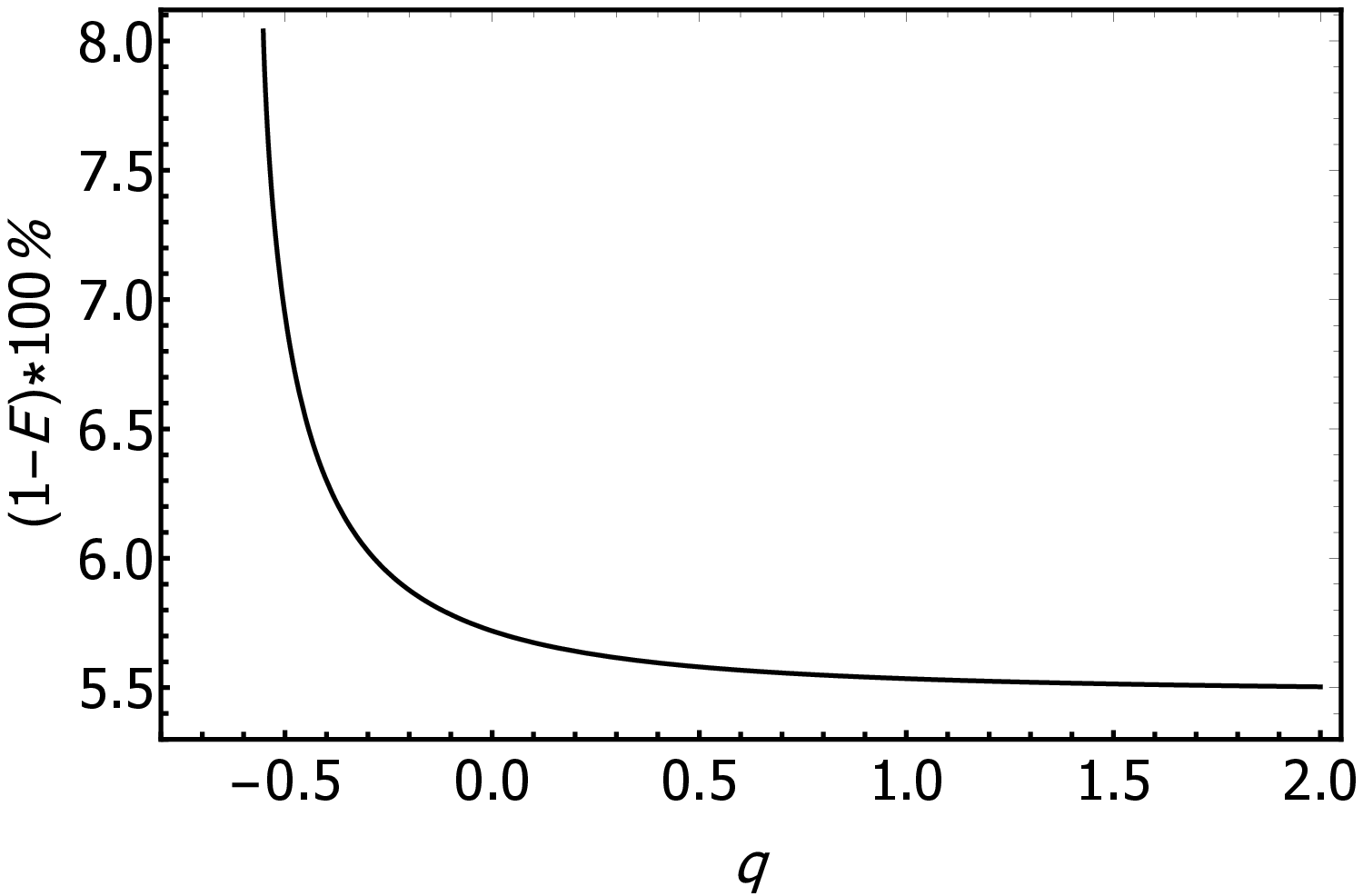} &
\includegraphics[width=0.48\linewidth]{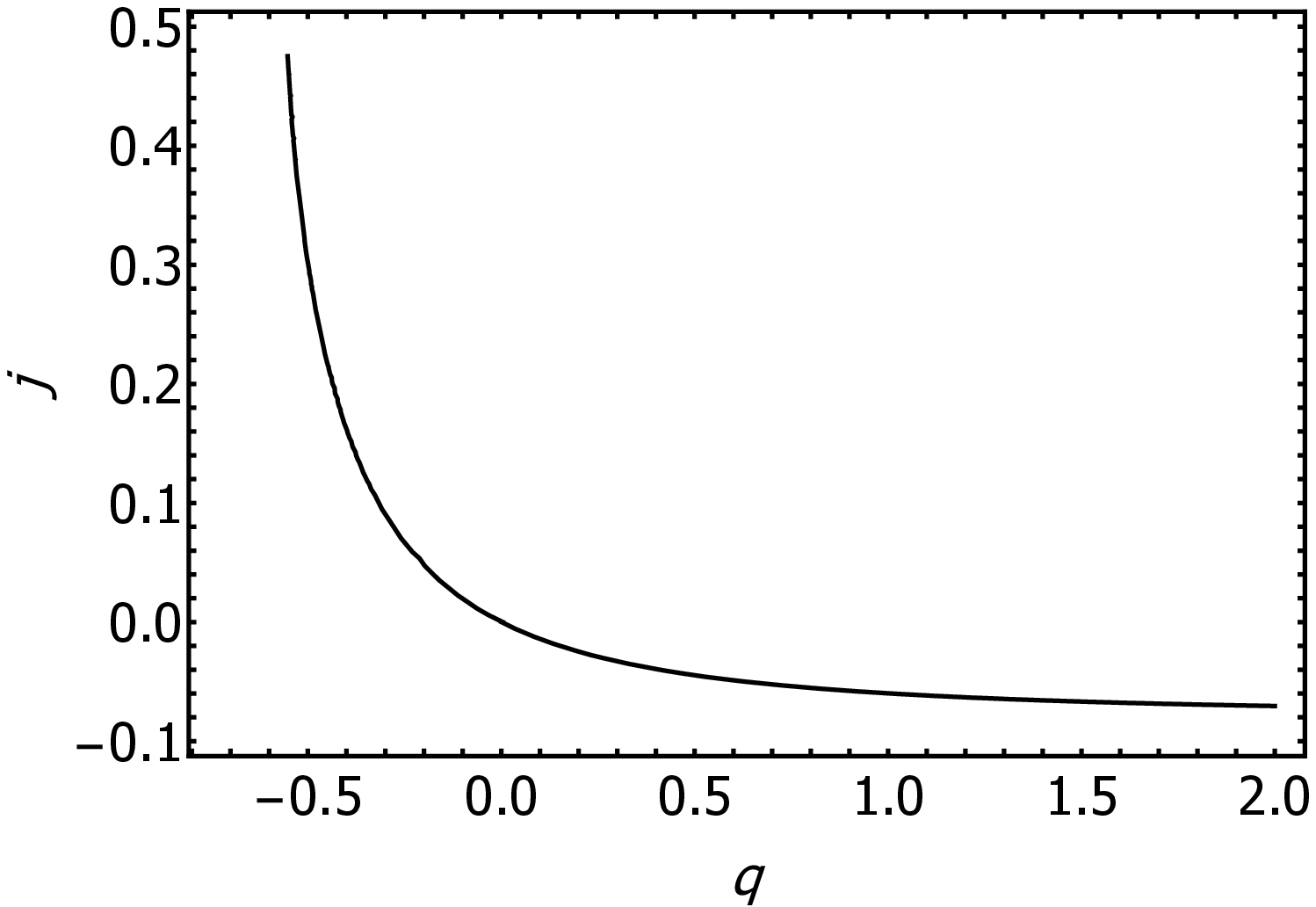} 
\end{tabular}
\caption{Left panel: Radiative efficiency for the oblate $q$-metric. Notice that the oblate $q$-metric is less efficient in converting accreting mass into radiation with respect to Schwarzschild. Right panel: Degeneracy between the radiative efficiency in the $q$-metric and the Kerr metric. For each value of the deformation parameter $q$ there exist a value of the angular momentum $j$ in the Kerr space-time for which the accretion disks in two geometries have the same efficiency.}
\label{fig:eff}
\end{figure*}

Finally in Fig.~\ref{fig:eff} we calculate the efficiency of accretion disks in the $q$-metric to convert mas into radiation. We notice that oblate sources ($q>0$) are less efficient than Schwarzschild, while prolate sources ($q<0$) are more efficient (left panel). Also we can see that there always exist a value of $q>0$ for which an oblate static source can produce the same efficiency as that of a Kerr black hole with counter-rotating disk (right panel).

\section{Final outlooks and perspectives}\label{sez5}

We considered the $q$-metric, a static and axially symmetric vacuum solution of Einstein's equations, as the possible exterior field of an exotic compact object and considered the eventuality that such a source may be distinguished from a Kerr black hole from the observation of the accretion disk's spectrum. 

To do so, we derived the orbital parameters of test particles in the equatorial plane of the $q$-metric for oblate (i.e. positive $q$) sources, calculated $r_{ISCO}/M_T$ and obtained the expression of measurable quantities such as the radiative flux, differential luminosity and efficiency of the accretion disk. 

We showed that the appearance of oblate sources may be mimicked by a Kerr space-time with counter-rotating disk ($j<0$). However, the flux emitted by the $q$-metric appears to be larger at very small radii (i.e. close to the ISCO) with respect to Schwarzschild and Kerr with negative $j$, thus suggesting the possibility that the spectral features of the two kinds of accretion disks may be distinguished.

Further, under the assumption of black body emission from the disk, we constructed the spectral luminosity as a function of the frequency of emitted radiation in the $q$-metric and compared it with the one in the Kerr metric.  Whereas for the the Kerr metric we see the luminosity increasing (decreasing) with respect to Schwarzschild for all the frequencies with $j>0$ ($j<0$), the $q$-metric shows a different behavior with the luminosity being larger (smaller) at smaller (larger) frequencies. This suggesting the possibility that spectral features of accretion disks around compact objects may be used to distinguish the two geometries.

Of course the features of the spectra emitted by real accretion disks surrounding compact objects in the universe are much more complicated than the simple toy models employed here. For this reason such simple models can not be used to practically determine the actual geometry in the proximity of astrophysical black hole candidates. However, we believe that the general considerations obtained here do indicate a road towards the possibility in the future of experimentally constraining geometric quantities, such as mass quadrupole moment, of compact sources and thus answer the question whether astrophysical black holes are indeed described by the Kerr metric.

\begin{acknowledgments}
KB, TK, EK and OL acknowledge the Ministry of Education and Science of the Republic of Kazakhstan, Grant: IRN AP08052311. 
\end{acknowledgments}


\end{document}